\documentclass[aps,pra,amsfonts,superscriptaddress]{revtex4}
\usepackage{bbm} 
\usepackage{multirow}
\usepackage{epsfig}

\newcommand{\eq}{\begin{eqnarray}} 
\newcommand{\en}{\end{eqnarray}}

\def\ket#1{\mathinner{|{#1}\rangle}}
\newcommand{\braket}[2]{\langle #1|#2\rangle}

\begin{document}

\title{Quantum statistical synchronization of non-interacting particles}
\author{Malte C. Tichy}
\address{Physikalisches Institut, Albert--Ludwigs--Universit\"at Freiburg, Hermann--Herder--Strasse~3, D--79104 Freiburg, Germany }
\author{Markus Tiersch }
\address{Physikalisches Institut, Albert--Ludwigs--Universit\"at Freiburg, Hermann--Herder--Strasse~3, D--79104 Freiburg, Germany }
\address{Institute for Quantum Optics and Quantum Information, Austrian Academy of Sciences, Technikerstrasse 21A, A--6020 Innsbruck, Austria }
\author{Fernando de Melo}
\address{Physikalisches Institut, Albert--Ludwigs--Universit\"at Freiburg, Hermann--Herder--Strasse~3, D--79104 Freiburg, Germany }
\address{Instituut voor Theoretische Fysica, Katholieke Universiteit Leuven, Celestijnenlaan 200D, B--3001 Heverlee, Belgium }
\author{Florian Mintert}
\address{Physikalisches Institut, Albert--Ludwigs--Universit\"at Freiburg, Hermann--Herder--Strasse~3, D--79104 Freiburg, Germany }
\address{Freiburg Institute for Advanced Studies, Albert-Ludwigs-Universit\"at Freiburg, Albertstrasse 19, D-79104 Freiburg, Germany}
\author{Andreas Buchleitner}
\address{Physikalisches Institut, Albert--Ludwigs--Universit\"at Freiburg, Hermann--Herder--Strasse~3, D--79104 Freiburg, Germany }

\begin{abstract}
A full treatment for the scattering of an arbitrary number of bosons through a Bell multiport beam splitter is presented that includes all possible output arrangements. Due to exchange symmetry, the event statistics differs dramatically from the classical case in which the realization probabilities are given by combinatorics. A law for the suppression of output configurations is derived and shown to apply for the majority of all possible arrangements. Such multiparticle interference effects dominate at the level of single transition amplitudes, while a generic bosonic signature can be observed when the average number of occupied ports or the typical number of particles per port is considered. The results allow to classify in a common approach several recent experiments and theoretical studies and disclose many accessible quantum statistical effects involving many particles.
\end{abstract}

\maketitle

\section{Introduction}
Non-interacting, distinguishable particles exhibit independent and therefore uncorrelated behavior. Due to the bosonic or fermionic nature of identical particles, however, such statement is no longer true for indistinguishable particles, even if no interaction takes place. For example, the bosonic nature of photons is impressively demonstrated by their statistical behavior in a  Hong-Ou-Mandel (HOM) setup \cite{Hong:1987mz}. Here, two identical photons are sent simultaneously (within their coherence time) through the two input ports of a balanced beam splitter. Due to the lack of interaction between the photons, one would not expect any correlations in the number of photons measured at both output ports. For fully indistinguishable photons, however, the particles always leave the setup together, and never exit at different ports. 

Such synchronization of two non-interacting particles has lead to many applications in quantum information sciences. The visibility of the HOM-dip quantifies the indistinguishability of two photons \cite{Ou:2006ta}. Thereby, the quality of single-photon sources can be tested \cite{Sun:2009dk}. The maximally entangled $\ket{\Psi^-}$ Bell-state (in any degree of freedom carried by the photons) can be detected or created, since it leads to an unambiguous signature in the setup \cite{PhysRevLett.61.2921}. This projection onto an entangled state can be applied in entanglement swapping protocols \cite{Halder:2007th} and quantum metrology \cite{Walther:2004it}. 

It is therefore of great interest to generalize the HOM setup for more than two photons and more than two input or output ports, i.e. to $n$ particles that are scattered in a setup with $n$ input and output ports. This would allow applications such as entanglement swapping or entanglement detection for many particles and the experimentally controlled transition from indistinguishability to distinguishability for many identical particles \cite{Tichy:2009kl,prepa}. 

While a comprehensive understanding of this scattering scenario is not yet available due to the complexity of the problem and the prohibitive scaling of the number of output states, several steps have been undertaken in this direction. The measurement of the  enhancement of events with all particles in one port - bunching events, was realized experimentally \cite{Ou:1999rr,Niu:2009pr}, a prediction for the suppression of coincident events for a specially designed biased setup with three particles and three input ports was presented \cite{Campos:2000yf}. The case of a Bell multiport beam splitter \cite{PhysRevA.55.2564,PhysRevA.71.013809} which redistributes $n$ incoming particles to $n$ ports in an unbiased way was discussed in \cite{Lim:2005qt}, where it was shown that coincident events are suppressed when $n$ is even. 

In this contribution, we extend our recent results \cite{Tichy:2010kx} on the characterization of the probabilities of \emph{all} possible output events of the Bell multiport beam splitter when $n$ particles are prepared in the $n$ input ports. Such treatment enables a general understanding of multiparticle interference effects, as well as on the average behavior of bosons. It hence unifies previous experimental and theoretical work on multiport beam splitters, and opens up new perspectives for the experimental verification and exploitation of bosonic multiparticle behavior. 

\section{Formalism}
\subsection{Setup and notation}
\begin{figure}[h]
\center
\includegraphics[width=9.25cm,angle=0]{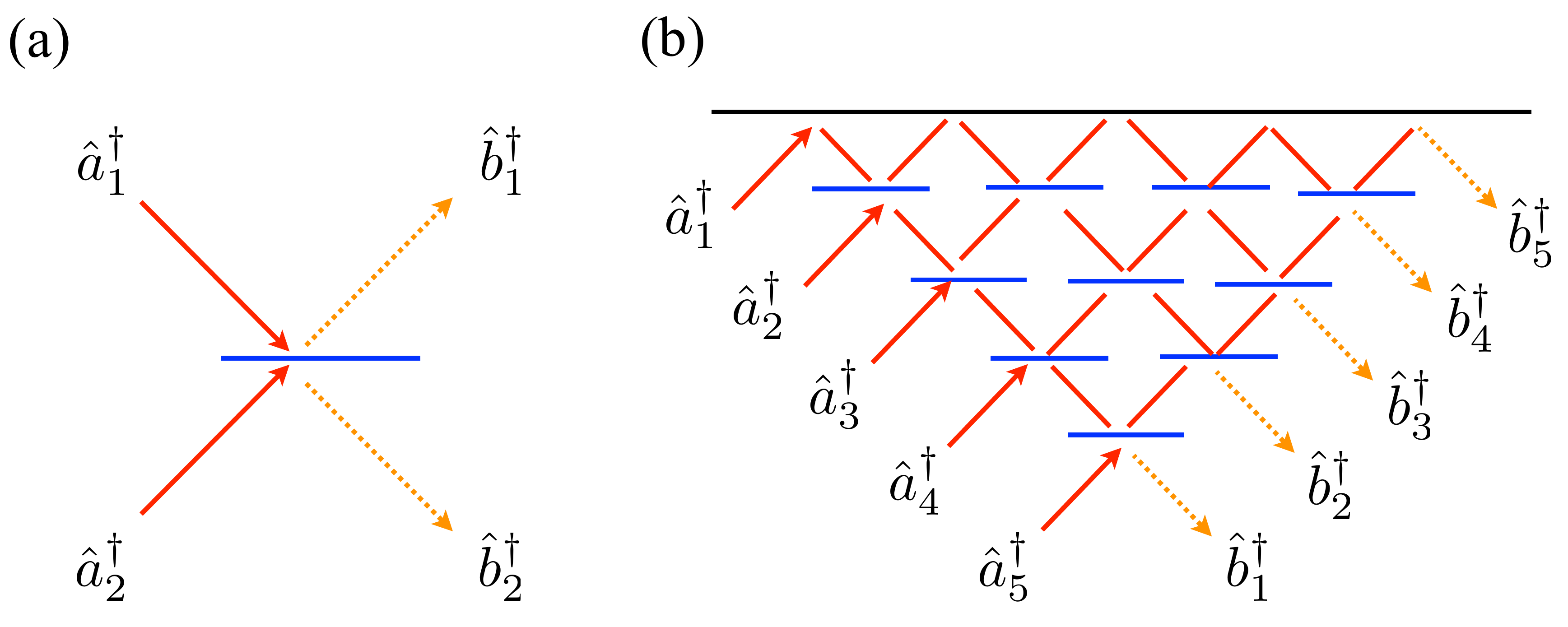} 
\caption{(a) Two-photon HOM scattering setup with two photons impinging on a balanced beam splitter. (b) Generalization: multiport beam splitter, here with five incoming and outgoing modes. } \label{setup}
\end{figure}

We consider a scattering scenario in which $n$ particles are initially evenly distributed among $n$ input modes. They are scattered by the multiport beam splitter and exit among $n$ output ports. The probability for each particle to exit at any port is the same, i.e. $1/n$. Such setup can be realized for photons via a simple beam splitter in the two-photon HOM case, as illustrated in Fig. (\ref{setup}a). A pyramidal combination of beam splitters with different reflection/transmission rates yields the generalization for $n$ ports and particles, such setup is shown in Fig. (\ref{setup}b), for $n=5$. 

We denote arrangements of $n$ particles in the $n$ modes by a vector $\vec s=(s_1, s_2, \dots s_n)$, with $s_k$ the number of particles in the output mode $k$, and  $\sum_{i=1}^n s_i = n$. Alternatively, we define the \emph{port assignment vector} $\vec d$ of length $n$ with entries that specify each particle's output port.  
It is constructed by concatenating $s_j$ times the port number $j$: \eq \vec d=\oplus_{j=1}^n \oplus_{k=1}^{s_j} (j) , \label{drepres} \en  e.g., for the arrangement $\vec s=(2,1,0,2,0)$, we find $\vec d(\vec s)=(1,1,2,4,4)$. 
\subsection{Distinguishable particles}
For distinguishable particles, no many-particle interference takes place, and the probability for a certain arrangement $\vec s$ is given by simple combinatorics: \eq P_{\mbox{class}}(\vec s) = \frac {1}{n^n} \frac{n!}{\prod_{j=1}^n s_j!}  \label{classprob} \en 
Due to the lack of interference phenomena, we call this situation ``classical''. In accordance to our intuition,  probabilities are summed instead of amplitudes. Two classes of events are especially interesting due to their extremal character. \emph{Coincident events}, i.e. $\vec s_c=(1,1,\ldots 1)$, are realized with probability $n!/n^n$. \emph{Bunching events}, with all particles at one output mode $k$, correspond to $s_k=n$ and thus to $\vec s_b=(0,0,..,n,..0)$. They are realized with probability $1/n^n$ and suppressed by a factor of $n!$ with respect to the coincident events. For large $n$, both events are highly unlikely, extreme cases.

\subsection{Indistinguishable bosons}
We reformulate the scattering scenario for identical particles, in second quantization. Applications of our study are feasible with today's optical technologies  \cite{PhysRevA.55.2564}, therefore, we focus on bosons. 
The initial state with one particle in each input port reads  $  \ket{\Psi}= \prod_{i=1}^n \hat a_i^\dagger \ket 0 \label{input} .$ The single-particle unitary evolution induced by the scattering setup acts on all particles independently and maps the input port creation operators $\hat a^\dagger_i$ to output creation operators  $\hat b_i^\dagger$ via a unitary matrix $U$ \cite{PhysRevA.40.1371}, such that $ \hat b_j^\dagger = \sum_{k=1}^n U_{jk} \hat a_k^\dagger . $
The unbiased Bell multiport beam splitter under consideration here corresponds to the unitary operation given by the Fourier matrix, defined for any dimension  $n$ by \eq U_{jk}=\frac{e^{\frac{2 \pi i}{n} (j-1)(k-1)}}{\sqrt{n}} . \label{fourier} \en  
The possible states with fixed particle number per port after the scattering process read  
\eq  \ket{ \Phi(\vec s)} = \left( \prod_{j=1}^n \frac 1 {\sqrt{s_j!}}\left(\hat  b_j^\dagger \right)^{s_j} \right) \ket 0. \en The transition probability to a specific output arrangement $\vec s$  can be written with the help of the port assignment vector (Eq. \ref{drepres}) as
\eq P_{\mbox{qm}}(\vec s)  = |\braket{\Psi}{\Phi(\vec s)}|^2= \frac{1}{  \prod_j s_j!  } \left| \sum_{\sigma \in P_n}  \prod_{j=1}^n  U_{d_j(\vec s),\sigma(j)} \right|^2  , \label{thebigsum}  \en
where $P_n$ denotes the set of all permutations of $\left\{1,..,n\right\}$. This coherent sum over $n!$ terms expresses the interference that occurs between all many-particle amplitudes that lead to the same output state. 
\subsection{Equivalence classes}
In order to discuss the behavior of the scattering system, it is necessary to identify classes of final states that occur with equal probability, within the quantum and the classical case. In the latter case, the realization probability of any arrangement $\vec s$, Eq. (\ref{classprob}), remains invariant under permutation of the output ports $s_k$. Hence we can define \emph{classical} equivalence classes which identify arrangements related to each other by permutation.
Ultimately, however, all final events $\vec s$ have to be considered as inequivalent.  In the case that the scattering matrix is a Fourier matrix such as given in Eq. (\ref{fourier}), some symmetry properties allow to reduce considerably the number of equivalence classes. One indeed finds that the amplitude (\ref{thebigsum}) is invariant under cyclic and anticyclic permutations. This allows us to define a \emph{quantum} equivalence relation between arrangements, and $N_{\mbox{quant}}$ associated quantum equivalence classes. 

The number of classical equivalence classes corresponds to the partition number, i.e. the number of possibilities to write an integer as sum of positive integers, while the total number of inequivalent events grows much faster, it is given by $N_{\mbox{total}}=\frac {(2n)!}{2 (n!)^2}$.  For comparison, the number of equivalence classes are given in Table \ref{suppressedd}.

\section{Event-suppression law}
In general, the evaluation of the transition probabilities in Eq. (\ref{thebigsum}) is a difficult task and cannot be performed in polynomial time with $n$ \cite{Lim:2005bf}. It is, however, possible to exploit the 
symmetry of the matrix $U$ to formulate a powerful law which predicts the suppression of final events. 
Indeed, since only $n$-th roots of unity appear in the Fourier matrix (Eq. \ref{fourier}), also every term of the $n!$ summands in Eq. (\ref{thebigsum}) can be written as such. 
Thereby, Eq. (\ref{thebigsum}) turns into \eq \braket{\Psi}{\Phi(\vec s)}= \sum_{k=0}^{n-1} c_k e^{i \frac{2 \pi}{n} k}, \en where the $c_k$ are natural numbers which give the cardinality of the following sets, defined in analogy to \cite{Graham:1976nx}, 
\eq u_r(\vec s) &=& \left\{ \sigma \left| \Theta_{n,\vec s}(\sigma) \equiv  \sum_{l=1}^n d(\vec s)_l \sigma(l) \right. = r \mbox{ mod } n\right\}, \en  with $ c_r = |u_r(\vec s)|$. 
The sum corresponds to the position of the barycenter of the set of points $\{ c_k e^{i \frac{2 \pi}{n} k} | k \in \{1,..,n \} \}$ in the complex plane. We set $Q=\mbox{mod}\left(\sum_{l=1}^{n} d_l(\vec s),n\right)$, and 
define an operation $\gamma$ which acts on permutations such that $\gamma(\sigma)(k) = \sigma(k) + 1 \mbox{ mod } n $. We find that $ \Theta_{n, \vec s}\left( \gamma(\sigma) \right) = \Theta_{n,\vec s} \left( \sigma \right) + Q. $
Thus, if $Q\neq 0$, the repeated application of $\gamma$ gives us a bijection between all pairs of $u_{r+a \cdot Q}$, for $a \in \left\{0,1,..,n-1 \right\}$. Hence, we find \eq \forall r\in \{0,..,n-1\}, \forall a \in \mathbbm{N}: c_{r+a \cdot Q} =c_{r} .\en If $Q\neq 0$, the set of points $\{c_k e^{i \frac{2\pi}{n} k }| k \in \{1,..,n \}  \}$ describes several interlaced polygons centered at the origin, ensuring that the sum vanishes, hence the process with the final state $\vec s$ is suppressed in this case. Thus, without knowing the values of the individual $c_k$, and only by symmetry properties, it is possible to predict that the total sum (Eq. \ref{thebigsum}) vanishes. This observation allows us to formulate: 
\eq Q(\vec s) := \mbox{Mod}\Big( \sum_{l=1}^n d_l(\vec s) ,n \Big)  \neq 0 \ \Rightarrow \ \braket{\Psi}{\Phi(\vec s)}=0 \label{ourtheorem} .\en

The law can be applied on any final state in an efficient way: consider, e.g., $n=6$ and $\vec s_1=(2,1,2,1,0,0)$. The port assignment vector reads $(1,1,2,3,3,4)$, and one finds $Q(\vec s_1)=2$, and this event is hence strictly suppressed. Unexpectedly though, the event $\vec s_2=(0,1,2,0,2,1)$, which is obtained from $\vec s_1$ by simple permutation, gives $Q(\vec s_2)=0$ due to the different port assignment vector $(2,3,3,5,5,6)$. It is actually enhanced by a factor larger than seven as compared to the classical event probability (also see Table \ref{fullvanishingratio}).

\subsection{Suppressed arrangements}
It is possible to estimate the number of suppressed arrangements predicted with the help of (\ref{ourtheorem}) by a simple argument. Since the number of arrangement $N_{\mbox{quant}}$ is much larger than $n$, we can  assume that the $Q(\vec s)$ are uniformly distributed in the interval $[0,\dots,n-1]$ for the ensemble of events $\vec s$. Then the probability to find a suppressed arrangement is given by the weight of nonvanishing values of $Q(\vec s)$, i.e., by $(n-1)/n=1-1/n$. This estimate is also numerically confirmed in the values shown in Table \ref{suppressedd}.  
\begin{table*}[t]
\begin{minipage}{.6\linewidth} 
\begin{tabular}{crrrrr}
$n$ & $N_{\mbox{total}}$ &$N_{\mbox{class}}$  & $N_{\mbox{quantum}}$ & $N_{\mbox{law}}$ & $N_{\mbox{supp}}$ \\ \hline
2 & 3&2& 2 & 1 & 0\\ 
3 & 10&3&  3& 1 & 0\\ 
4&  35&5&   8 &  5 & 0\\ 
5&  126&7 &  16 & 10 & 0 \\ 
6&  462& 11 &  50 & 38 & 2 \\ 
7&  1716&15 &  133 & 105 & 0 \\
8&  6435&22 &  440 & 371 & 0 \\ 
9&  24310&30 &  1387 & 1201 & 0\\ 
10& 92378& 42&     4752& 4226 & 96\\ 
11& 352716&56&    16159& 14575 &0\\ 
12  &1352078& 77&    56822& 51890 & 1133 	\\ 
13 & 5200300&101 &  200474 & 184626& 0 \\ 
14 & 20058300&135 &  718146 & 666114 & 2403 \\ 
\end{tabular} 
 \caption{Total number of events ($N_{\mbox{total}}$), classical equivalence classes ($N_{\mbox{class}}$), quantum equivalence classes ($N_{\mbox{quantum}}$), classes that therewithin fulfill the law \ref{ourtheorem} ($N_{\mbox{law}}$), and suppressed classes which are not predicted by Eq. \ref{ourtheorem} ($N_{\mbox{supp}}$). }
	  	 \label{suppressedd}	 
  \end{minipage}
    \hspace{.02\linewidth}
\begin{minipage}{.4\linewidth} 
\begin{tabular}{clc}
$n$ & $\vec s$ & Enhancement \\ \hline 
 3 & (003) & 6 \\ 
  & (111) & 3/2 \\ \hline  
 4 &   (0004) & 24 \\ 
  &   (0202) , (0121)&  8/9  \\ \hline  
$ 5$ &   (00005)  & 120 \\ 
 &   (00131),  (01103)  & 15/2 \\ 
 &  (00212),  (01022)  & 10/3 \\ 
 &  (11111)   & 5/24\\ \hline 
$ 6$ & $ (000006) $ & 720 \\ 
& (002004), (000141),  & \multirow{3}{*}{\Bigg\} 144/5 \hspace{1.2mm} } \\ 
 &(010104),  (000303),&  \\
 & (001032), (000222)  &\\
	   & (020202), (001113),  & \multirow{2}{*}{\Big\} 36/5 \hspace{1.64mm} } \\ &  (012021)  \\ 
	 \end{tabular} 
 \caption{ Nonsuppressed output states, together with the corresponding quantum enhancement, i.e., the ratio of quantum to classical event probability.} 
	 \label{fullvanishingratio}
  \end{minipage}
\end{table*}
\begin{table}[t]
\end{table}
 
\subsection{Application of the suppression law}
\begin{figure}[h]
\center
\includegraphics[width=10.5cm,angle=0]{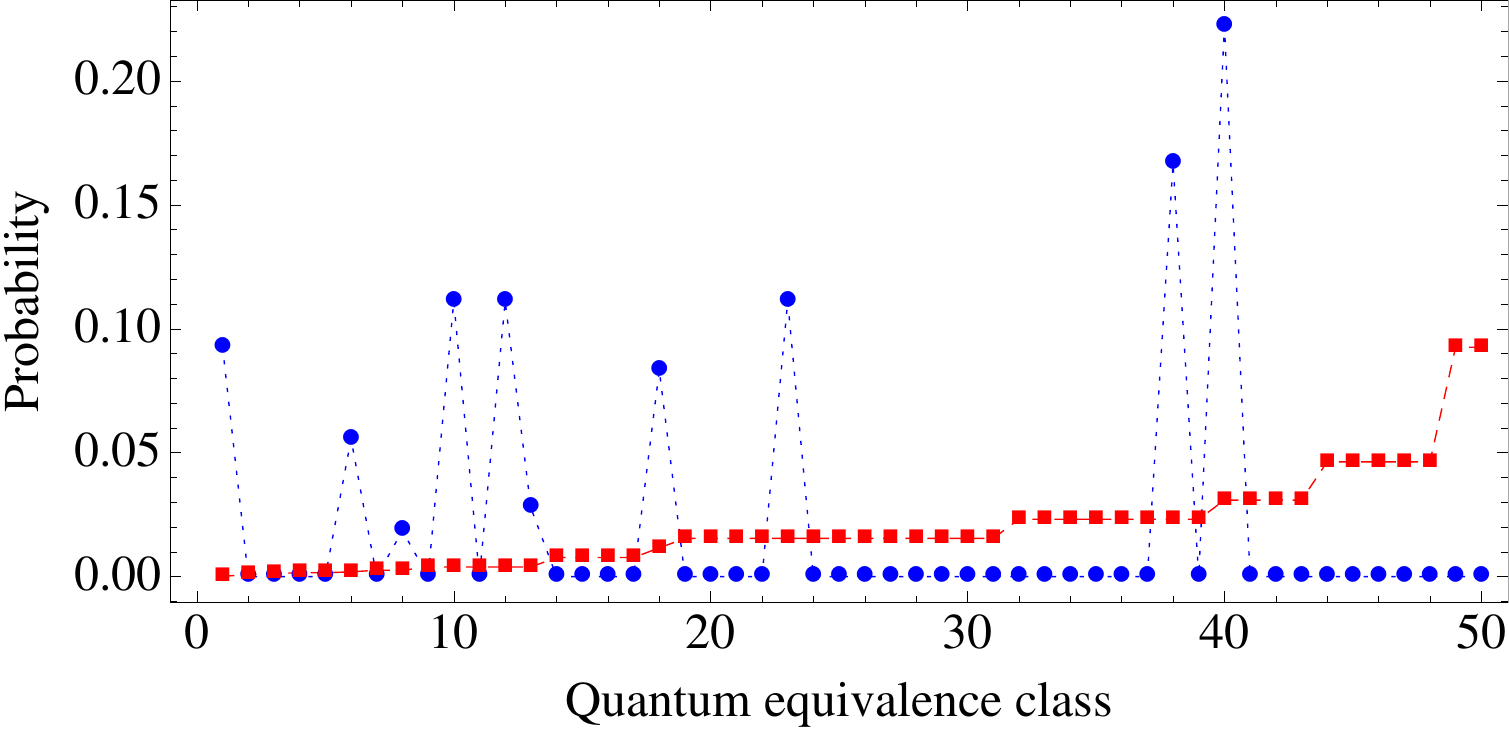} 
\caption{Quantum (blue circles) and classical (red squares) probability for the realization of the 50 different quantum equivalence classes for $n=6$ particles, sorted by their classical realization probability. Note that most states are fully suppressed in the quantum case, only a few are highly enhanced with respect to the classical case. } \label{sixp}
\end{figure}

\begin{figure}[h] \center
\includegraphics[width=5.5cm,angle=0]{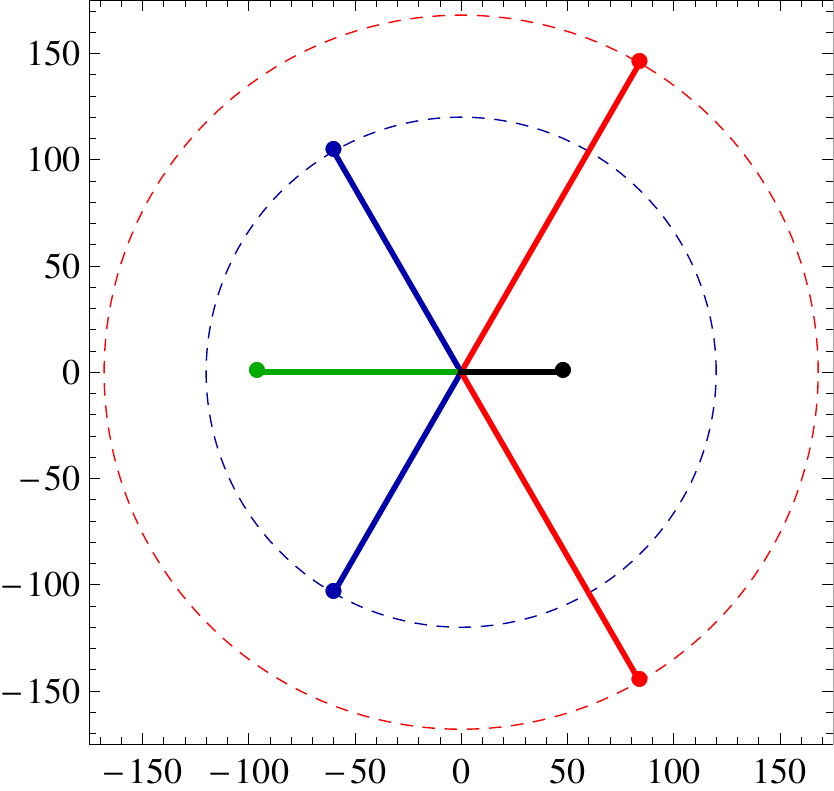}  \hspace{.02\linewidth}
 \includegraphics[width=5.5cm,angle=0]{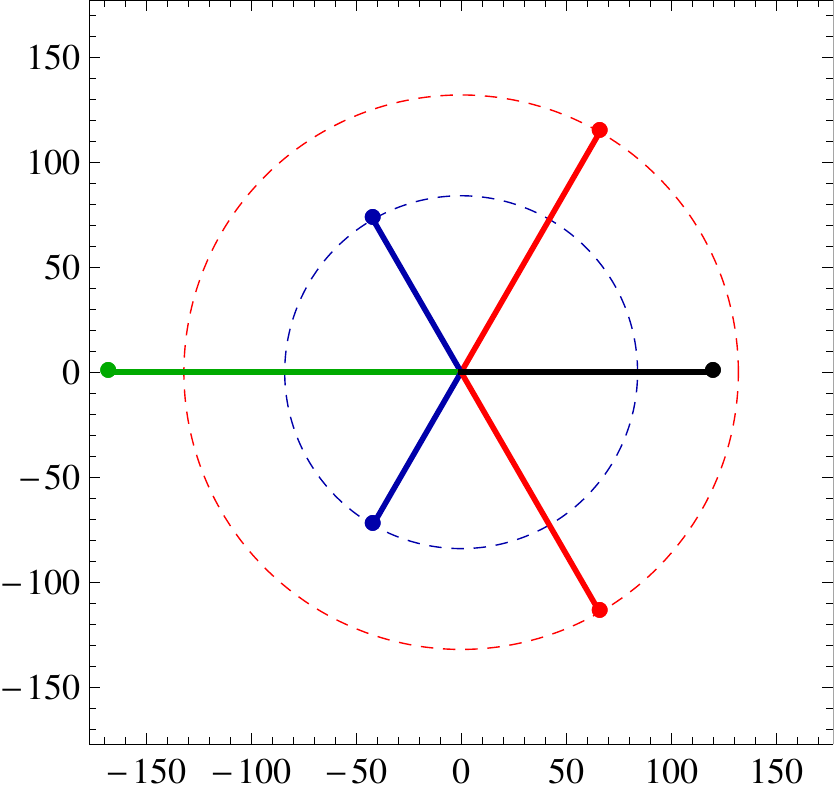} 
\caption{Illustration of the corresponding $c_k$ in the complex plane for the process with $\vec s=(0, 1, 2, 1, 0, 2)$ (left hand side) and $\vec s=(0, 1, 1, 2, 1, 1)$ (right hand side). The sum of the complex numbers vanishes while the points do not lie all on circles.} \label{supp6}
\end{figure}

For $n=2..6$, we list the unsuppressed arrangements in Table \ref{fullvanishingratio}, together with their \emph{quantum enhancement}, i.e., the ratio of quantum-to-classical event probability. From the results, it is apparent that the behavior of the system becomes more extreme, the more particles are involved: the enhancement factor is bounded from above by $n!$, a value that is reached for the enhancement for bunching events with $\vec s=(n,0,..,0)$. For the 50 quantum equivalent arrangements that exist for $n=6$, we show the classical and quantum probabilities in Fig. \ref{sixp}. Two arrangements are suppressed although they do not fulfill the requirements of the law: $\vec s=(0,1,2,1,0,2)$ and $\vec s=(0,1,1,2,1,1)$. In Fig. \ref{supp6}, we show the values of the corresponding $c_k$ in the complex plane. One can easily see that while the values do not lie on polygons, the sum of all contributions still vanishes. Such situations are exceptional, as can be seen from Table \ref{suppressedd}.

\section{Bosonic behavior}
As we have seen in the last section, the implications of Eq.~(\ref{ourtheorem}) on the realization probabilities of single events are important for the overall behavior of the system: most events are totally suppressed, while only few remain which are highly enhanced. Intuitively, one would expect that events with many particles in one port are generally favored by bosons. Indeed, bunching events are always enhanced by a factor of $n!$ with respect to the classical case. The number of particles in one port or the number of occupied ports does, however, turn out not to be a good indicator for the enhancement or the suppression of a certain event. For example, events of the type $\vec s=(n-1,1,0,..,0)$ could be expected to be enhanced due to the bosonic nature of the particles, while they actually turn out to be strictly suppressed, for all $n$. Thus, at the level of the event probabilities of single arrangements, interference effects dominate, and the bosonic nature of the particles is not apparent at all. 

Such general bosonic behavior is recovered when a coarse-grained grouping of many final arrangements in larger classes is performed. Such classes can be characterized, e.g., by the number of occupied ports $k$,  by the number $m$ of particles in one port, or by the classical equivalence classes. The event probability for such a class is given by the sum of the probabilities of the single events that pertain to the class. 

\begin{figure}[h]
\center
\includegraphics[width=10.5cm,angle=0]{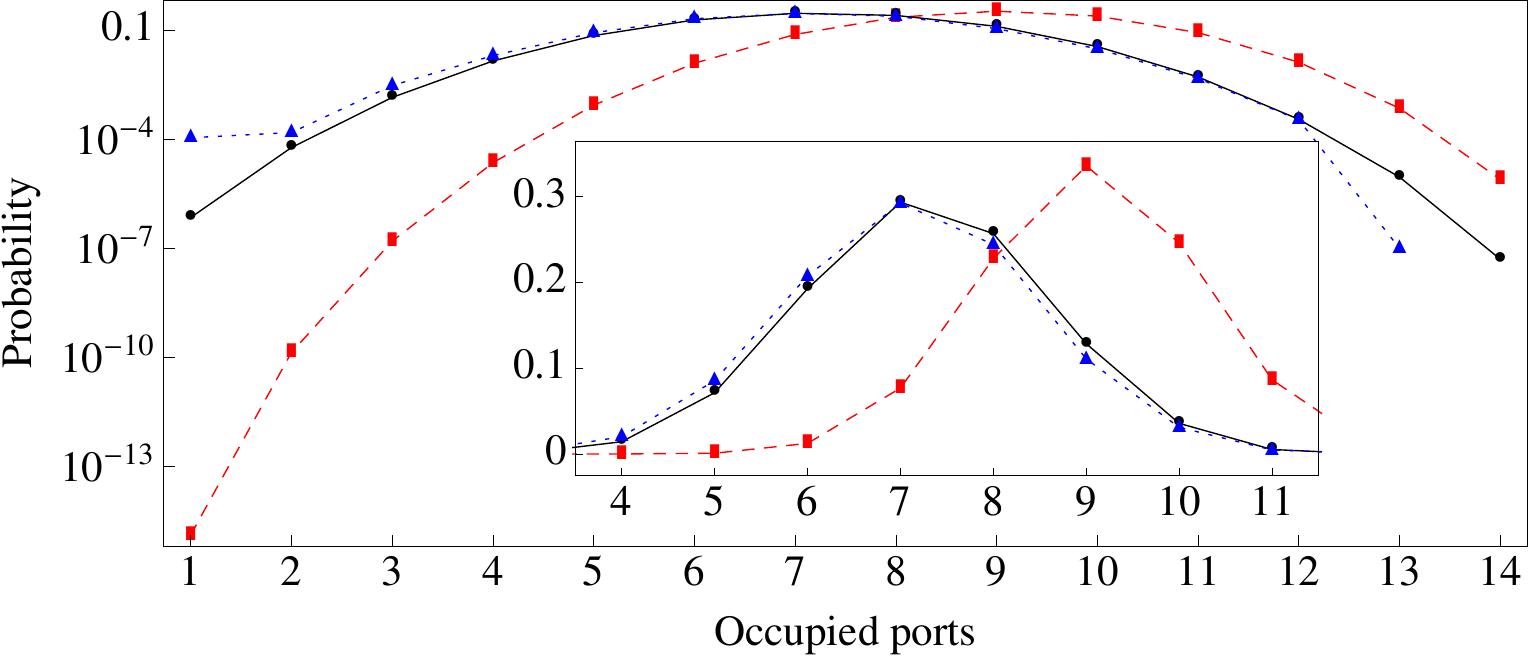} 
\caption{Event probability for a given number of occupied ports, for $n=14$. Red squares denote classical combinatorics, blue triangles the quantum calculation, and black circles our (bosonic) estimate for the quantum result. The inset shows the same distribution on a linear scale. Events with 14 occupied ports are strictly suppressed in the quantum case.}\label{nosingle2}
\end{figure}

When performing such average, we expect that interference effects disappear while the bosonic enhancement of states with many particles in one port persists. This can be also seen in our formalism: according to (\ref{thebigsum}), the probabilities $P_{qm}(\vec s)$ are given in terms of a sum over permutations of \emph{scattering amplitudes}, i.e., over complex numbers of equal modulus (products of matrix-elements of $U_{jk}$). Since these numbers typically have different phases, they tend to add up destructively. However, all $s_j!$ permutations $\sigma$ that interchange the $s_j$ particles that exit in port $j$ leave the scattering amplitudes invariant, so that $s_j!$ terms in the sum have equal phases and add up constructively. This motivates the following approximation for the transition probability (\ref{thebigsum}):
\eq P_{\mbox{approx}}(\vec s) = \frac{\left( \prod_j s_j! \right) P_{\mbox{class}}(\vec s)}{\sum_{\vec r} \left( \prod_j r_j! \right) P_{\mbox{class}}(\vec r)} . \label{semiesti} \en 
We show the probability distribution for the number of occupied ports, for the classical calculation (Eq. \ref{classprob}), for the bosonic quantum case (Eq. \ref{thebigsum}), and for our approximation (Eq. \ref{semiesti}), for $n=14$, in Fig. \ref{nosingle2}.  As expected, bosons always tend to occupy less output ports than in the classical case. This behavior is persistent for any $n$. Furthermore, Fig. \ref{nosingle2} shows that the approximation (Eq. \ref{semiesti}) predicts the actual outcome very well for most $k$, and only fails for events with almost all or almost no sites occupied. This is easily understood, since, for very small or very large $k$, few distinct equivalence classes contribute to these event groups. Then, again interference dominates the event probability, rather than bosonic behavior. 

\begin{figure}[h]
\center
\includegraphics[width=10.5cm,angle=0]{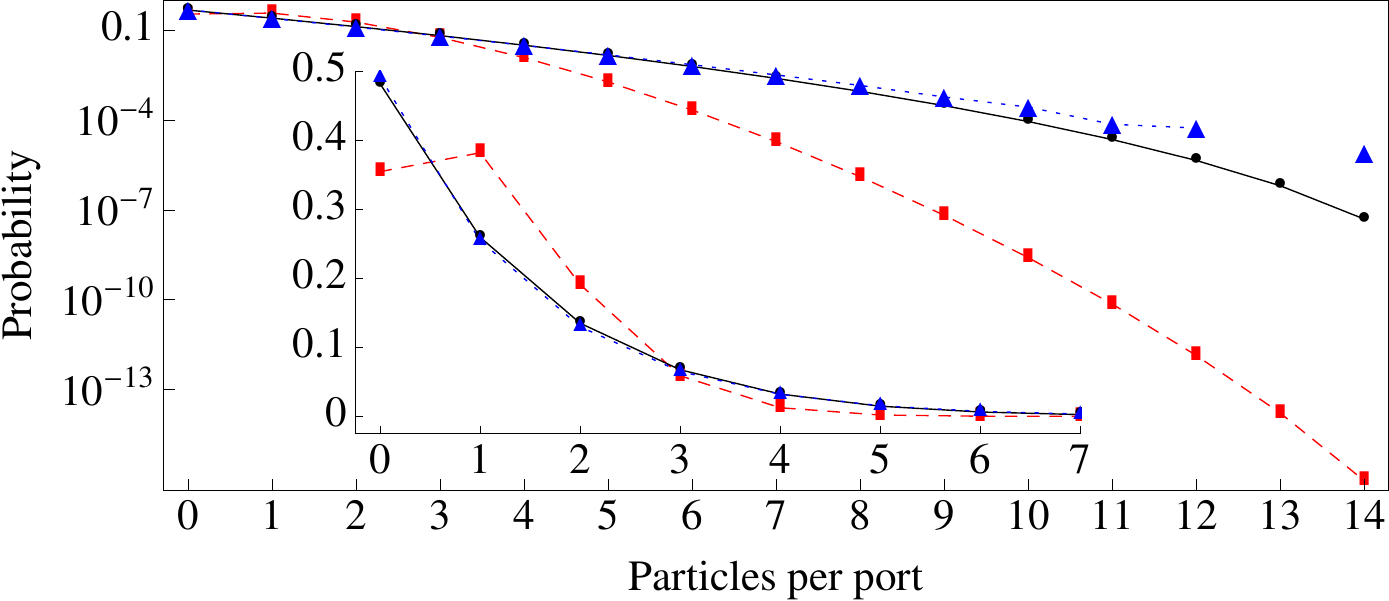} 
\caption{Probability to find exactly $k$ particles (horizontal axis) in one port, for $n=14$. Red squares denote the classical, blue triangles the quantum calculation, and black circles the estimate (\ref{semiesti}). The inset shows the distribution on a linear scale for small $k$. Events with 13 particles in one port are totally suppressed for bosons. }\label{occudistri}
\end{figure}

Also the event probability for a given number of particles in one single port is well described by our estimate (Eq. \ref{semiesti}). 
For 14 particles, the probability distribution is shown in Fig. \ref{occudistri}. Again, we see a dramatic difference between the classical and quantum case, especially for the probability to find a large number of particles in one port. 

\begin{figure}[h]
\center
\includegraphics[width=11.5cm,angle=0]{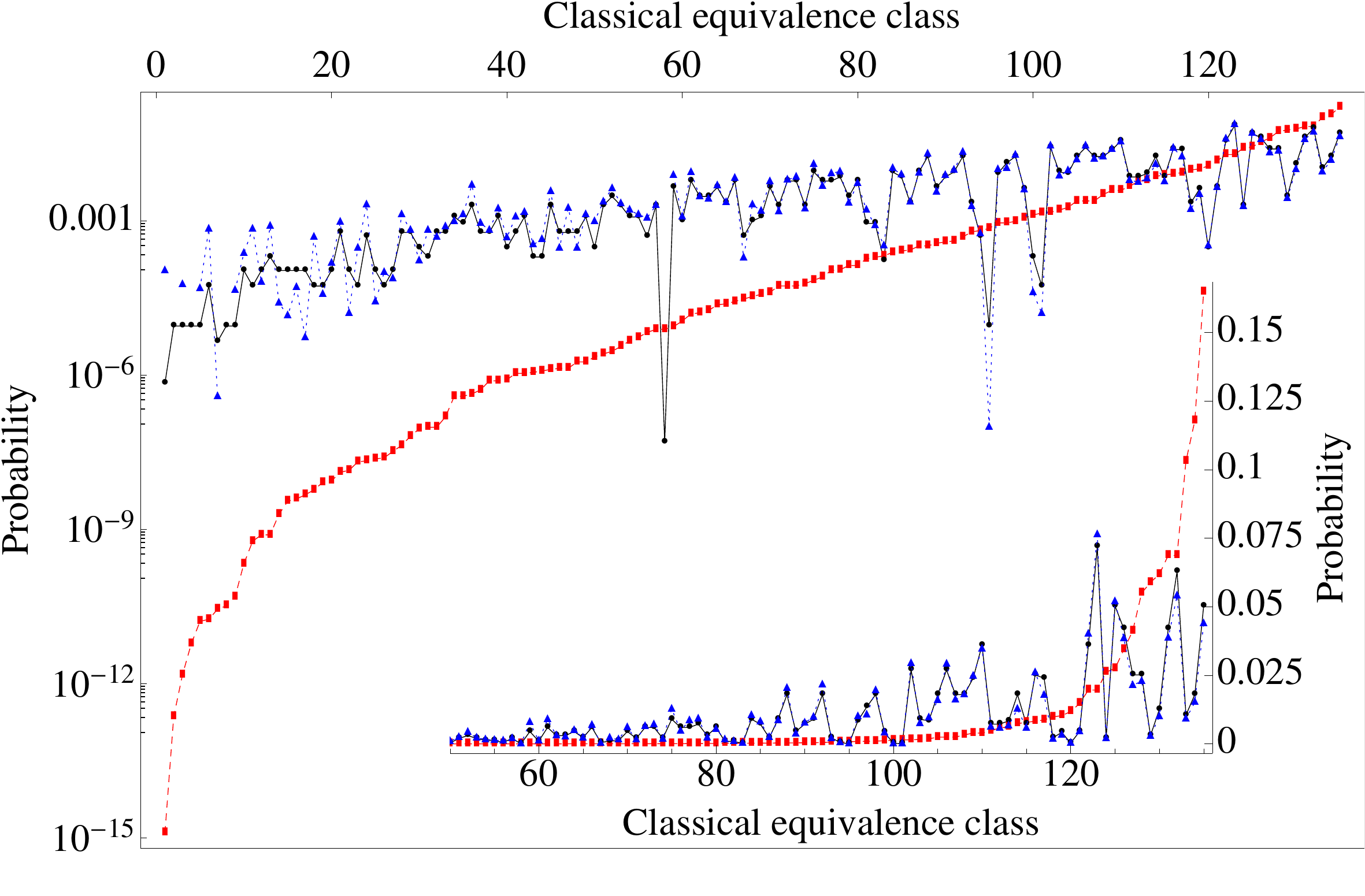} 
\caption{Quantum (Blue triangles) and classical probability (Red rectangles) and estimate (Black circles) for the realization of events grouped according to the 135 classical equivalence classes. The classical equivalence classes are ordered according the classical realization probability given by Eq. (\ref{classprob}). The classical probability is therefore -- by construction -- a monotonically increasing function. The inset shows a detail of the probability distribution in linear scale.} \label{classclasses}
\end{figure}

A further application of the bosonic approximation can be performed when we group the 718146 events according to their 135 classical equivalence classes. The resulting probabilities are shown in Fig. \ref{classclasses}. While this grouping is much more fine than in the previous two examples, the difference between the classical and quantum case is still very well pronounced and reproduced by the estimate.

\section{Conclusions and outlook}
Synchronization of non-interacting particles might seem contradictory by construction. However, it turns out to be possible due to the exploitation of quantum statistical effects stemming from the indistinguishability of particles. We generalized the most prominent example for such a behavior, the HOM effect, to $n$ particles and $n$ ports on two different levels: interference effects inhibit the realization of most possible events for single transition amplitudes, while general statistical characteristics with smooth bosonic behavior emerge which are efficiently approximated by Eq. (\ref{semiesti}). On the fine as well as on the coarse grained scale, however, quantum and classical transmission probabilities differ dramatically.
\section*{Acknowledgements}
M.C.T. acknowledges financial support by Studien\-stiftung des deutschen Volkes, F.d.M. by the Belgium Interuniversity Attraction Poles Programme P6/02, and F.M. by DFG grant MI 1345/2-1, respectively.

\end{document}